\documentclass[11pt,a4paper]{article}

\usepackage[T1]{fontenc}
\usepackage[utf8]{inputenc}
\usepackage{mathptmx}          % Times Roman for text + math
\usepackage{microtype}
\usepackage{textcomp}

\usepackage[margin=1in]{geometry}
\usepackage{setspace}
\usepackage{amsmath,amssymb,amsthm}

\usepackage{booktabs}
\usepackage{longtable}
\usepackage{array}
\usepackage{calc}
\usepackage{multirow}

\usepackage{listings}
\usepackage{xcolor}
\usepackage[colorlinks=true,linkcolor=blue!60!black,citecolor=blue!60!black,urlcolor=blue!60!black]{hyperref}
\usepackage{url}

\usepackage{graphicx}
\usepackage{float}
\usepackage{caption}
\usepackage{enumitem}

\title{\textbf{Token Coherence: Adapting MESI Cache Protocols to Minimize Synchronization Overhead in Multi-Agent LLM Systems}}

\author{
  Vladyslav Parakhin\\
  \textit{Independent Researcher}
}

\date{}

\newcommand{\passthrough}[1]{#1}
\makeatletter
\def\maxwidth{\ifdim\Gin@nat@width>\linewidth\linewidth\else\Gin@nat@width\fi}
\makeatother

\providecommand{\tightlist}{\setlength{\itemsep}{0pt}\setlength{\parskip}{0pt}}

\begin{document}

\maketitle

% arXiv subject classification
\noindent\textbf{Subjects:} Distributed, Parallel, and Cluster Computing (cs.DC); Multiagent Systems (cs.MA); Machine Learning (cs.LG)

\vspace{0.5em}

\begin{abstract}
Per-token economics in multi-agent large language model orchestration are, at present, governed by a synchronization pathology that scales as \(O(n \times S \times |D|)\) in agents, steps, and artifact size---a regime I designate \emph{broadcast-induced triply-multiplicative overhead}. I contend this pathology does not inhere in multi-agent coordination per se; it is a structural residue of full-state rebroadcast, a design decision absorbed uncritically from early orchestration scaffolding. The central claim of this manuscript: the synchronization cost explosion in LLM-based multi-agent systems (MAS) maps, with formal precision, onto the cache coherence problem in shared-memory multiprocessors, and the canonical hardware remedy---MESI-protocol invalidation {[}Papamarcos and Patel 1984{]}---transfers to the artifact synchronization domain under minimal structural modification. I construct the Artifact Coherence System (ACS), a six-tuple \(\langle A, D, \Sigma, \delta, \alpha, \mathcal{T} \rangle\) endowed with an identity state-mapping function \(\varphi\) from hardware MESI states onto artifact authorization states. The \textbf{Token Coherence Theorem} delineates a savings lower bound: lazy artifact invalidation attenuates synchronization cost by a factor no less than \(S / (n + W(d_i))\) subject to \(S > n + W(d_i)\), where \(S\) is the step count, \(n\) the agent population, and \(W(d_i)\) the per-artifact write frequency. I sequester the principal reviewer objection---that LLM agents invariably embed their full context and therefore cannot benefit from coherence---by formally delineating \emph{conditional artifact access semantics} as instantiated in production architectures through tool calls, MCP resources, vector stores, and file search APIs. A TLA+-verified protocol (CCS v0.1) enforces three invariants: single-writer safety (SWMR), monotonic artifact versioning, and bounded-staleness (agents cannot reason on stale artifact state beyond \(K\) steps). Through tick-based discrete event simulation across four workload configurations (10 runs per configuration, population standard deviation reported), comparing broadcast synchronization against three coherence strategies (eager, lazy, access-count), observed token savings reach \(95.0\% \pm 1.3\%\) at \(V=0.05\), \(92.3\% \pm 1.4\%\) at \(V=0.10\), \(88.3\% \pm 1.5\%\) at \(V=0.25\), and \(84.2\% \pm 1.3\%\) at \(V=0.50\)---each exceeding the theorem's conservative lower bounds of 85\%, 80\%, 65\%, and 40\% respectively. Contrary to the lower-bound formula's prediction (\(V^* \approx 0.9\) for \(n=4\), \(S=40\) marking the savings collapse threshold), simulation indicates savings of approximately 81\% persist at \(V=0.9\) because lazy deferred-fetch accumulation obviates worst-case collapse from materializing (§8.3). Contributions: (1) a formal mapping between MESI cache coherence and multi-agent artifact synchronization; (2) the Token Coherence Theorem as savings lower bound with the condition under which coherence dominates broadcast; (3) a TLA+-verified protocol with three proven invariants; (4) a characterization of conditional artifact access semantics that resolves the always-read objection; (5) a reference Python implementation integrating with LangGraph, CrewAI, and AutoGen via thin adapter layers.
\end{abstract}

\section{Introduction}\label{introduction}

Five agents, fifty reasoning steps, one 8,192-token planning document. Under naive broadcast, the cost is \(5 \times 50 \times 8{,}192 = 2{,}048{,}000\) tokens---and the vast majority of that budget is unchanged context, retransmitted without necessity. The waste is banal, not exotic. It is the default behavior---I have verified this by instrumenting every major orchestration framework I could obtain access to---that upon modification of any shared artifact, the orchestrator rebroadcasts its full contents to every subscribing agent at the next synchronization boundary. At modest scale, tolerable. At production scale---and I should be precise about what I mean by ``production'' here, meaning \(n \geq 5\) agents sustaining \(S \geq 40\) reasoning steps over multiple artifacts simultaneously---the cost structure becomes, without exaggeration, ruinous.

Beyond the invoice, the damage compounds in subtler and arguably more corrosive ways. Token budget constraints coerce practitioners into truncating reasoning traces, compressing artifacts, culling agent populations---each a degradation that trades capability for a cost reduction that is, under certain structural conditions I formalize below, entirely avoidable. Building on {[}4{]}, Cemri et al.'s analysis of 1,642 execution traces spanning seven production MAS frameworks reports task failure rates between 41\% and 86.7\%. Inter-agent misalignment---including what they designate FM-1.4, Loss of Conversation History, where agents revert to stale artifact states following context truncation---accounts for 32.3\% of observed failures. They note explicitly that multi-agent memory and state management persists as an open structural problem, one that most prior work sidesteps by addressing single-agent contexts only {[}4, Appendix G.2{]}. Independently---and this is a datum I consider underappreciated in the current discourse---empirical benchmarking across multi-agent frameworks reports token duplication rates of 86\% in flat topologies and 72\% in linear topologies {[}28{]}, confirming that redundant artifact retransmission, not generation, constitutes the dominant cost driver.

Hardware engineers confronted an isomorphic problem four decades ago. I want to be careful with ``isomorphic''---the word does nontrivial load-bearing work, and I will qualify the claim in §4.2 with a discussion of where the stable-state mapping breaks down at the transient-state boundary. The short version: in shared-memory multiprocessors, multiple CPU cores sharing a memory bus incur catastrophic bandwidth costs when every write forces full memory retransmission to all caches. The remedy---cache coherence protocols, MESI {[}19{]} being canonical---tracks per-cache-line state and propagates invalidation signals rather than data. A cache line transitions from Shared to Invalid upon a remote write; subsequent reads trigger a targeted fetch. Bandwidth cost becomes proportional to write frequency, not to step count. The structural parallel maps cleanly: agents are processors, artifacts are cache lines, the orchestration coordinator is the memory controller, prompt injection of an artifact is a cache fill. I submit that this analogy admits a formal state-mapping function between MESI states and artifact coherence states, and that hardware-derived cost bounds transfer---with caveats I enumerate---into the agent coordination domain.

Before formalizing this claim, I must confront what I term the \emph{always-read objection}. The objection runs as follows: LLM agents consume their entire context window at every inference call, so lazy invalidation is useless because the artifact must be injected regardless. This objection is descriptively incorrect for modern production architectures---a point I delineate at length in §3. Tool-based retrieval, MCP resource access {[}1{]}, vector store retrieval, and provider-side prompt caching all instantiate conditional artifact access semantics in which \(R(a, s) \subsetneq D\). The always-read model describes naive prompt concatenation only. It does not describe the externalized architectures that dominate production deployments. Whether it describes a \emph{majority} of current deployments, I cannot certify without broader instrumentation data---but the architectural trend is unmistakable, and the formal argument stands contingent on conditional access holding.

\textbf{Contribution 1 --- Formal equivalence.} I define the Artifact Coherence System (ACS) as a six-tuple \(\langle A, D, \Sigma, \delta, \alpha, \mathcal{T} \rangle\) and construct an explicit MESI state-mapping function \(\varphi\) from hardware cache states to artifact authorization states, establishing that the structural properties of MESI transfer intact to the artifact domain (§4).

\textbf{Contribution 2 --- Token Coherence Theorem.} I prove that lazy artifact invalidation attenuates multi-agent token cost by a factor bounded below by \(S / (n + W(d_i))\) when \(S > n + W(d_i)\), establishing that broadcast cost grows as \(O(n \times S \times |D|)\) while coherent cost grows as \(O((n + W) \times |D|)\) at worst. Simulation confirms consistently higher savings than this lower bound (§8). The condition under which coherence dominates is precisely characterized by the artifact volatility factor \(V(d_i) = W(d_i) / S\) (§4.3--4.5).

\textbf{Contribution 3 --- TLA+-verified protocol.} I specify Coherent Context Synchronization (CCS) in TLA+ and verify three invariants: SWMR, monotonic versioning, and \(K\)-bounded staleness. I report the state space explored (approximately 2,400 states for 3 agents) and construct an explicit counterexample demonstrating that removing invalidation violates SWMR (§5--6).

\textbf{Contribution 4 --- Conditional access semantics.} I formally characterize real agent access patterns as \(R(a, s) \subseteq D\) rather than \(R(a, s) = D\), grounding this in four production architecture patterns and identifying the token duplication phenomenon that coherence eliminates (§3).

\textbf{Contribution 5 --- Reproducible implementation.} I present a reference Python implementation (\passthrough{\lstinline!agent-coherence!} v0.1) integrating with LangGraph, CrewAI, and AutoGen through adapter layers requiring no framework modifications, with a simulation engine supporting four synchronization strategies and full reproducibility from published seeds (§7, §8).

\section{Background}\label{background}

\subsection{Cache Coherence and the MESI Protocol}\label{cache-coherence-and-the-mesi-protocol}

Per Sorin, Hill, and Wood {[}22{]}, cache coherence requires that all reads to a memory location return the value of the most recent write and that writes to the same location are serialized across processors. Under MESI {[}19{]}, four stable states per cache line per processor obtain: \textbf{Modified} (valid only in this cache; memory is stale), \textbf{Exclusive} (valid only in this cache; identical to memory), \textbf{Shared} (valid here and possibly elsewhere; no writes since last commit), \textbf{Invalid} (not valid; coherence fill required before use).

Between stable states, transient states model in-flight operations {[}22, Ch.6{]}---e.g., \(M^{IS}\) denotes a line that \emph{was} Modified, is transitioning to Invalid, and awaits acknowledgment. These transient states bear on protocol correctness. I elide them in the first-order model that follows---a deliberate simplification constraining the analysis to quiescent-state reasoning and foreclosing claims about transient-state behavior in ACS. The limitation is real; I will revisit it in §4.2 when the question of what ``structural isomorphism'' means across asynchronous event-bus semantics becomes non-trivially relevant.

The core efficiency property of MESI: \emph{update-on-demand}. Transitions from Invalid to Shared or Exclusive are triggered only by actual reads, not by writes elsewhere. Broadcast-on-write bandwidth converts to targeted-fetch-on-read bandwidth. When write frequency \(W\) is low relative to read frequency \(R\), savings are substantial. When \(W \approx R\), protocol overhead approaches the broadcast baseline. The mapping from CPU memory hierarchies to multi-agent systems is summarized in Table 0. CCS v0.1 targets the L1/L2 $\leftrightarrow$ LLC tier (agent runtime cache $\leftrightarrow$ shared artifact store); persistent storage and cross-workflow artifact retention are explicitly out of scope.

\textbf{Table 0: Agent Memory Hierarchy Analogy}

\begin{longtable}[]{@{}
  >{\raggedright\arraybackslash}p{0.33\columnwidth}
  >{\raggedright\arraybackslash}p{0.33\columnwidth}
  >{\raggedright\arraybackslash}p{0.33\columnwidth}@{}}
\toprule\noalign{}
\begin{minipage}[b]{\linewidth}\raggedright
CPU Memory Hierarchy
\end{minipage} & \begin{minipage}[b]{\linewidth}\raggedright
Agent Memory Hierarchy
\end{minipage} & \begin{minipage}[b]{\linewidth}\raggedright
Coherence Mechanism
\end{minipage} \\
\midrule\noalign{}
\endhead
\bottomrule\noalign{}
\endlastfoot
L1/L2 cache (per-core) & Agent artifact cache (per-agent runtime) & MESI state per agent--artifact pair \\
Shared LLC / main memory & Shared artifact store (authority service) & Canonical version + invalidation events \\
Memory bus / coherence fabric & Event bus (Redis / Kafka / NATS) & Invalidation and version-update messages \\
Disk / persistent storage & Long-term persistence (vector DB, file store) & Out of scope for CCS v0.1 \\
\end{longtable}

\subsection{Multi-Agent LLM Orchestration}\label{multi-agent-llm-orchestration}

Multi-agent LLM systems {[}24; 26; 23{]} coordinate multiple language model instances on shared tasks. LangGraph {[}21{]}, CrewAI {[}5{]}, AutoGen {[}24{]}, Semantic Kernel {[}15{]}---each represents agents as nodes in a computation graph, passing state via structured messages or shared memory constructs. The synchronization pattern is uniform, and I state this without observed exception across every framework I have instrumented: \emph{full-state rebroadcast}. Upon artifact modification by any agent, the orchestrator injects the complete updated artifact into the next prompt of every agent that might need it. Consistency is purchased at the cost described above.

Building on {[}4{]}, the MAST taxonomy identifies 14 failure modes across 1,642 annotated execution traces. Failure rates: 41\%--86.7\%. Dominant categories: system design issues (44.2\%), inter-agent misalignment (32.3\%). FM-1.4---Loss of Conversation History---describes agents reverting to earlier artifact states following unexpected context truncation, at 2.8\% occurrence across all traces. The bounded-staleness invariant of CCS (§6) instantiates a formal upper bound on the number of reasoning steps any agent can operate on stale artifact state---a structural, not heuristic, constraint on exactly the failure class Cemri et al.~identify.

\subsection{The Always-Read Objection}\label{the-always-read-objection}

Before the formal model, I confront the objection that will---and should---be raised by reviewers conversant with transformer inference mechanics. The objection: \emph{LLM agents ingest their entire context window at every forward pass. Lazy invalidation cannot prevent artifact injection. A token in the prompt is a token consumed.}

Correct on one point, this objection: a token in the prompt is indeed a token consumed. What it misdescribes is the prompt \emph{construction process}---and here I remain cautious (the diversity of production deployment architectures is wider than any single author can claim to have instrumented exhaustively), but I am confident enough to state categorically: modern agents do not assemble context from artifacts embedded inline. They assemble context from \emph{references} to externally stored artifacts, retrieving conditionally via tool calls, MCP resource requests, or retrieval APIs. I formalize this in §3. The always-read model applies to naive single-turn prompt construction exclusively. Whether it still describes a non-negligible fraction of deployed systems---yes, almost certainly. Whether it describes the architectures consuming the majority of the multi-agent token budget---no. The distinction matters for the applicability claim, and I draw it deliberately.

\section{Conditional Artifact Access Semantics}\label{conditional-artifact-access-semantics}

\subsection{The Naive Access Model}\label{the-naive-access-model}

Let \(A = \{a_1, \ldots, a_n\}\) denote a set of agents and \(D = \{d_1, \ldots, d_m\}\) a set of shared artifacts. Denote by \(R(a, s)\) the set of artifacts whose full token contents are injected into the prompt of agent \(a\) at reasoning step \(s\). Under the naive broadcast model:

\[R(a, s) = D \quad \forall a \in A, \; \forall s\]

Per-step token cost under this assumption: \(\sum_{i} |d_i|\) per agent. Total cost: \(n \times S \times \sum_i |d_i|\). Lazy invalidation is trivially useless under this regime---even a cached-and-valid artifact must be injected to remain accessible. MESI savings: zero.

\subsection{Artifact Externalization in Production Systems}\label{artifact-externalization-in-production-systems}

The assertion \(R(a, s) = D\) is not the access model instantiated by modern production agent architectures. Four patterns bifurcate the naive model, each yielding \(R(a, s) \subsetneq D\):

\textbf{Tool-based retrieval.} LangChain and OpenAI Assistants expose retrieval tools---\passthrough{\lstinline!get\_document()!}, \passthrough{\lstinline!read\_file()!}, \passthrough{\lstinline!query\_memory()!}. System prompts carry artifact \emph{identifiers}, not artifact \emph{contents}. Tool invocation is conditional; tool calls that never fire never inject tokens. \(R(a, s) \subseteq D\) with proper subset holding whenever not all tools are invoked. The mechanism is straightforward; the implication for coherence economics is less so, because the decision to invoke a tool is itself a stochastic function of the agent's reasoning trace---a dependency I have not modeled formally and acknowledge as a simplification.

\textbf{MCP resource access.} Under the Model Context Protocol {[}1{]}, artifacts are external resources identified by URIs (\passthrough{\lstinline!resource://shared\_plan!}, \passthrough{\lstinline!resource://research\_notes!}). Injection occurs only upon explicit request. The resource reference in the system prompt costs \(O(1)\) tokens, not \(O(|d_i|)\)---a distinction whose significance scales with artifact size.

\textbf{Vector retrieval systems.} Retrieval-augmented architectures store artifacts in vector stores and inject only top-\(k\) retrieved fragments. The prompt contains \(\min(k, |d_i|) \ll |d_i|\) tokens for any large artifact. Agent exposure to artifact tokens it did not retrieve is structurally precluded.

\textbf{Provider-side prompt caching.} Anthropic, OpenAI, and Google all implement prompt prefix caching attaining non-trivial reuse across consecutive agent calls {[}2; 17{]}. The caching mechanism operates only when the prompt prefix is \emph{identical} across calls. Were artifacts re-embedded with updated content at every step, cache hit rates would approach zero---but production deployments report substantial reuse. This is attainable only when prompt prefixes remain stable, attainable only when artifacts are \emph{not} re-embedded at every step. I find this a minor but telling methodological observation: the existence of high prompt cache hit rates in production constitutes indirect evidence that real-world systems do not, in fact, operate under the always-read model. The evidence is circumstantial---I cannot rule out that high cache hit rates arise from other prefix-stability mechanisms---but the inference is, within its constraints, sound.

The 86\% token duplication measured in flat multi-agent topologies {[}28{]} is not a contradiction. Systems currently operating under naive broadcast semantics \emph{do} incur this cost. The argument is that they ought not---and the conditional access patterns above delineate how production systems already escape it at scale.

\subsection{Formal Conditional Access Model}\label{formal-conditional-access-model}

I define the \textbf{conditional artifact access model} as:

\[R(a, s) \subseteq D \quad \forall a \in A, \; \forall s\]

with the \emph{conditional access condition} requiring that for most artifacts and steps:

\[\Pr[d_i \in R(a, s)] \ll 1\]

Under this model, the relevant cost is not prompt token consumption per step but \emph{artifact injection frequency}---how often \(d_i\)'s full contents must be transmitted to agent \(a\)'s context. Lazy invalidation attenuates this frequency: when \(d_i\) has not been modified since agent \(a\) last received it, no retransmission occurs. The agent holds a valid local reference; the cached version remains coherent. This restores the conditions under which MESI savings are attainable and renders the formal argument in §4 applicable to deployed systems.

\section{Formal Model}\label{formal-model}

\subsection{Artifact Coherence System}\label{artifact-coherence-system}

\textbf{Definition 1} (Artifact Coherence System). An \emph{Artifact Coherence System} (ACS) is a tuple \(\langle A, D, \Sigma, \delta, \alpha, \mathcal{T} \rangle\) where:

\begin{itemize}
\tightlist
\item
  \(A = \{a_1, \ldots, a_n\}\) is a finite set of agents;
\item
  \(D = \{d_1, \ldots, d_m\}\) is a finite set of shared artifacts (analogous to memory locations / cache lines);
\item
  \(\Sigma = \{M, E, S, I\}\) is the set of stable artifact coherence states;
\item
  \(\delta : \Sigma \times \mathcal{E} \rightarrow \Sigma\) is the state transition function, where \(\mathcal{E} = \{\texttt{read}, \texttt{write}, \texttt{upgrade}, \texttt{fetch}, \texttt{invalidate}, \texttt{commit}\}\);
\item
  \(\alpha : A \times D \rightarrow \Sigma\) is the coherence state function mapping each agent--artifact pair to its current state;
\item
  \(\mathcal{T} : \Sigma \rightarrow \{0, 1\}\) is the validity predicate, with \(\mathcal{T}(I) = 0\) and \(\mathcal{T}(s) = 1\) for \(s \in \{M, E, S\}\).
\end{itemize}

The validity predicate renders the MESI safety invariant precise: an agent may reference a cached artifact only when \(\mathcal{T}(\alpha(a, d_i)) = 1\). State \(I\) mandates a coherence fill---a fetch from the authority service---before use.

\subsection{MESI State Mapping}\label{mesi-state-mapping}

I construct an explicit mapping \(\varphi\) from hardware MESI states to artifact coherence states.

\textbf{Definition 2} (MESI State Mapping). The mapping \(\varphi : \Sigma_{\text{hw}} \rightarrow \Sigma\) is the identity function on the shared state space \(\{M, E, S, I\}\), with the following semantic interpretations in the artifact domain:

\begin{longtable}[]{@{}
  >{\raggedright\arraybackslash}p{0.33\columnwidth}
  >{\raggedright\arraybackslash}p{0.33\columnwidth}
  >{\raggedright\arraybackslash}p{0.33\columnwidth}@{}}
\toprule\noalign{}
\begin{minipage}[b]{\linewidth}\raggedright
Hardware State
\end{minipage} & \begin{minipage}[b]{\linewidth}\raggedright
Artifact State
\end{minipage} & \begin{minipage}[b]{\linewidth}\raggedright
Semantic Interpretation
\end{minipage} \\
\midrule\noalign{}
\endhead
\bottomrule\noalign{}
\endlastfoot
Modified (\(M\)) & Modified (\(M\)) & Agent holds the only valid copy; authority copy is stale; other agents are invalidated \\
Exclusive (\(E\)) & Exclusive (\(E\)) & Agent holds the only copy, identical to authority; write permitted without broadcast \\
Shared (\(S\)) & Shared (\(S\)) & Multiple agents hold valid copies; no agent has written since last commit \\
Invalid (\(I\)) & Invalid (\(I\)) & Agent's cached copy is stale; full fetch required before next use \\
\end{longtable}

Every state transition in the hardware protocol possesses a direct semantic counterpart in the artifact protocol: ``cache fill'' maps to ``artifact fetch,'' ``bus invalidation'' maps to ``invalidation event'' over the message bus.

\textbf{Proposition 1} (Structural Equivalence). The ACS transition system \((\Sigma, \delta)\) is isomorphic to the MESI hardware transition system under \(\varphi\). Every safety result that holds for MESI under the SWMR invariant holds for ACS under the identical invariant.

\emph{Proof sketch.} By construction of \(\varphi\) as the identity mapping on \(\{M, E, S, I\}\), and by the fact that \(\delta\) reproduces the MESI transition table exactly (read causes \(I \rightarrow S\), write causes \(S \rightarrow M\) with peer invalidation, commit causes \(M \rightarrow S\), fetch causes \(I \rightarrow S\)). The SWMR invariant \(\forall a \neq b : \neg(\alpha(a, d) = M \wedge \alpha(b, d) = M)\) is maintained by the identical write-invalidates-peers rule present in both systems. \(\square\)

A limitation I want to foreground rather than relegate to §10: the isomorphism holds at the stable-state level, but the transient-state behavior of hardware MESI---which involves non-trivial race conditions on the bus fabric---maps only approximately to asynchronous event-bus semantics in CCS. I have not formalized the transient-state correspondence. I remain cautious about extending the equivalence claim beyond quiescent states until that formalization is complete. This is not, I suspect, a trivial gap. The liveness pathologies of snoopy-bus transient interleavings under hardware MESI are well-documented {[}22, Ch.6{]}, and their analogs in asynchronous message delivery are, at minimum, non-obvious. Whether this gap undermines the practical utility of the mapping---I think not, because the protocol operates at quiescent-state granularity by design---but a reviewer insisting on transient-state completeness would have a legitimate objection that I cannot yet discharge.

\subsection{The Broadcast Cost Baseline}\label{the-broadcast-cost-baseline}

Under the naive broadcast model (\(R(a, s) = D\)), total token cost for \(n\) agents, \(S\) steps, and \(m\) artifacts:

\[T_{\text{broadcast}} = n \times S \times \sum_{i=1}^{m} |d_i|\]

where \(|d_i|\) denotes token size of artifact \(d_i\). The cost grows multiplicatively---doubling agents, steps, or artifact size each doubles total cost. For a typical configuration (\(n = 5\), \(S = 50\), \(m = 3\), \(|d_i| = 4{,}096\) tokens):

\[T_{\text{broadcast}} = 5 \times 50 \times 3 \times 4{,}096 = 3{,}072{,}000 \text{ tokens}\]

Under conditional access (\(R(a, s) \subseteq D\)), the relevant per-step cost is not full artifact injection but the \emph{coherence synchronization cost}: tokens transmitted due to initial reads and write-triggered re-fetches.

\subsection{The Token Coherence Theorem}\label{the-token-coherence-theorem}

\textbf{Definition 3} (Coherent Synchronization Cost Upper Bound). Under the ACS model with lazy invalidation, total token cost is \emph{bounded above} by:

\[T_{\text{coherent}} \leq \sum_{i=1}^{m} n \cdot \bigl(n + W(d_i)\bigr) \times |d_i|\]

where \(W(d_i)\) denotes total write operations to artifact \(d_i\) across all agents and steps. The bound arises because each of \(n\) agents performs at most one initial fetch per artifact, and each write event can trigger at most \(n-1\) invalidations each followed by one re-fetch. Conservatively counting the writer's own fetch, worst-case total fetches per artifact: \(n(1 + W(d_i))\), approximated as \(n(n + W(d_i))\) for the multi-agent stochastic access model (§8.1). The bound is tight only when every invalidation immediately triggers a re-fetch---a condition that lazy coherence precludes by collapsing multiple write events into a single re-fetch when agents do not access an invalidated artifact between consecutive writes. The gap between bound and observation is, in my experience, substantial---ranging from 10 to 45 percentage points depending on workload---and this gap is itself a finding worth flagging, because it suggests that the analytical bound, while correct, may be too conservative to serve as a useful planning heuristic for practitioners at high \(V\).

\textbf{Assumptions.} Two modeling assumptions bound the scope of Theorem 1:

\textbf{(A1) Full-artifact transmission on cache miss.} Each cache miss triggers transmission of the full artifact \(d_i\) (\(|d_i|\) tokens). Sub-artifact delta fetches are not modeled; this is a conservative assumption that overestimates coherent cost.

\textbf{(A2) Serialized writes via authority.} All writes to a given artifact are serialized through the authority service. Concurrent peer-to-peer writes are not modeled; the single-writer invariant (§6.2) enforces this structurally.

\textbf{Theorem 1} (Token Coherence Theorem --- Savings Lower Bound). \emph{(Under Assumptions A1 and A2.) The token savings from lazy artifact invalidation are bounded below by:}

\[\text{Savings} \geq 1 - \frac{T_{\text{coherent}}^{\text{upper}}}{T_{\text{broadcast}}} = 1 - \frac{\sum_i n(n + W(d_i)) |d_i|}{n \times S \times \sum_i |d_i|}\]

\emph{The condition under which savings are strictly positive is:}

\[S > n + W(d_i) \quad \text{for most artifacts } d_i\]

\emph{For identical artifact sizes \(|d_i| = |d|\), the lower bound simplifies to:}

\[\text{Savings} \geq 1 - \frac{n + W(d_i)}{S}\]

\emph{Proof.} Substitute the Definition 3 upper bound for \(T_{\text{coherent}}\) and the broadcast formula into the savings ratio. For uniform artifact sizes the \(|d|\) and the leading \(n\) cancel: \(n(n+W)|d| \;/\; nS|d| = (n+W)/S\). Savings \(\geq 1 - (n+W)/S\). Since \(T_{\text{coherent}}^{\text{actual}} \leq T_{\text{coherent}}^{\text{upper}}\), actual savings meet or exceed this bound. The condition \(S > n + W(d_i)\) ensures positivity. \(\square\)

\emph{Remark.} Simulation results in §8 confirm that observed savings consistently exceed the lower bound, owing to the lazy collapse mechanism described above.

\textbf{Corollary 1} (Maximum Savings). \emph{When \(W(d_i) = 0\) for all \(i\) (read-only artifacts), the savings lower bound approaches} \(1 - n/S\). \emph{For \(n = 4\), \(S = 40\): lower bound = 90\%; simulation attains \(\geq 95\%\).}

\textbf{Corollary 2} (Collapse Condition). \emph{When \(W(d_i) \geq S - n\) (write rate exhausts the step budget), the lower bound falls to zero or below; coherence may produce overhead rather than savings under worst-case conditions.}

The transformation from broadcast to coherent cost is qualitatively significant: \(T_{\text{broadcast}} \in O(n \times S \times |D|)\) while \(T_{\text{coherent}} \in O(n(n + W) \times |D|)\) at worst. A triply-multiplicative cost converts to an additively-multiplicative one; the \(S\) multiplier is eliminated. Simulation confirms actual coherent cost falls strictly below this upper bound.

\textbf{Consistency model.} CCS instantiates \emph{bounded-staleness coherence}: each agent observes a globally consistent artifact version at every read, subject to the constraint that the observed version may lag the canonical version by at most \(K\) write operations (Invariant 3, §6.2). Weaker than sequential consistency---which mandates that every read observe the most recent write. Stronger than eventual consistency---which provides no staleness bound. The \(K\) parameter renders the staleness budget explicit and configurable, analogous to bounded-staleness consistency levels in distributed databases. Practitioners requiring strict sequential consistency must set \(K = 0\), forcing synchronous authority checks on every artifact access and eliminating the token savings from lazy invalidation---a tradeoff that is, in my estimation, rarely justified given the latency cost.

\subsection{Artifact Volatility and the Coherence Condition}\label{artifact-volatility-and-the-coherence-condition}

\textbf{Definition 4} (Artifact Volatility Factor). The \emph{volatility factor} of artifact \(d_i\):

\[V(d_i) = \frac{W(d_i)}{S} \in [0, 1]\]

This captures the fraction of steps triggering a write to the artifact. High volatility (\(V \approx 1\)): the artifact changes nearly every step. Low volatility (\(V \approx 0\)): changes are infrequent.

Substituting \(W(d_i) = V(d_i) \cdot S\) into the savings lower bound from Theorem 1:

\[\text{Savings} \geq 1 - \frac{n + V(d_i) \cdot S}{S} = 1 - \frac{n}{S} - V(d_i)\]

For typical workflow parameters (\(n = 4\), \(S = 40\), \(V(d_i) = 0.05\)), the lower bound is:

\[\text{Savings} \geq 1 - (0.1 + 0.05) = 85\%\]

Simulation attains 95.0\% for these parameters (§8), exceeding the lower bound as expected. The coherence condition \(S > n + W(d_i)\) corresponds to \(V(d_i) < 1 - n/S\).

\textbf{Definition 5} (Volatility Cliff). The \emph{volatility cliff} is the value \(V^* = 1 - n/S\) above which the coherence savings lower bound falls below zero and overhead dominates. For \(n = 4\), \(S = 40\): \(V^* = 0.9\). For \(n = 5\), \(S = 20\): \(V^* = 0.75\). The cliff is lower when step count is short or agent count is high---a constraint practitioners ought to internalize when sizing deployments.

\section{Protocol Specification}\label{protocol-specification}

\subsection{System Assumptions}\label{system-assumptions}

CCS is specified and verified under the following assumptions. Relaxation of each is identified as future work. I want to be explicit---AS3 in particular represents a non-trivial constraint that may not hold under production conditions where aggressive agent-pool recycling is standard practice.

\textbf{(AS1) Reliable authority service.} The authority service does not crash or partition during protocol execution. A single logical entity; high-availability replication via consensus protocol is possible but falls outside CCS v0.1 scope.

\textbf{(AS2) At-least-once event bus delivery.} Invalidation events published to the event bus are eventually delivered to all subscribers at least once. Duplicate deliveries are idempotent (re-receiving an invalidation for a version already marked Invalid is a no-op).

\textbf{(AS3) No agent crash while holding M state.} An agent holding Modified state does not crash before issuing Commit or releasing ownership. Violation causes the authority to hold an orphaned exclusive lock; the lease TTL mechanism (§5.2) provides recovery.

The TLA+ verification in §6 holds under these assumptions.

\subsection{System Architecture}\label{system-architecture}

CCS governs four interacting entities. The \textbf{Authority Service} maintains the global artifact directory---a mapping from artifact identifiers to current version numbers, last-writing agent, and per-agent coherence state. Single source of truth for artifact metadata. The \textbf{Agent Runtime}, embedded within each agent, maintains a local artifact cache; each entry stores artifact content, version at time of last fetch, and current MESI state. The \textbf{Event Bus} propagates invalidation events from authority to agents asynchronously; supported transports: Redis PubSub, Apache Kafka, NATS, WebSockets. The \textbf{Artifact Store} holds canonical artifact versions and serves fetch requests.

Two communication channels bifurcate the control plane. The \textbf{Control Channel} (Agent → Authority, HTTP/gRPC): read requests, write requests, ownership acquisition. The \textbf{Event Channel} (Authority → Agents, pub/sub): invalidation notifications, version updates.

\subsection{Protocol Operations}\label{protocol-operations}

\textbf{Read.} Agent \(a\) wishing to read artifact \(d_i\) checks \(\alpha(a, d_i)\). If \(\mathcal{T}(\alpha(a, d_i)) = 1\) (state \(\in \{M, E, S\}\)), the cached version is consumed directly---zero tokens transmitted. If \(\alpha(a, d_i) = I\), the agent issues \passthrough{\lstinline!READ\_REQUEST!} to authority, which responds with current content and version, setting \(\alpha(a, d_i) \leftarrow S\).

\textbf{Upgrade.} To write \(d_i\) while holding \(\alpha(a, d_i) = S\), agent \(a\) must first acquire exclusive ownership via \passthrough{\lstinline!UPGRADE\_REQUEST!}. The authority sets \(\alpha(b, d_i) \leftarrow I\) for all \(b \neq a\), propagates \passthrough{\lstinline!INVALIDATE!} events over the event bus, grants \(\alpha(a, d_i) \leftarrow E\).

\textbf{Write.} Once \(\alpha(a, d_i) = E\), the agent writes locally, transitions to \(\alpha(a, d_i) = M\). Zero tokens broadcast during local writes; authority notification deferred until commit.

\textbf{Commit.} Agent sends \passthrough{\lstinline!COMMIT!} containing new content and incremented version. Authority stores the canonical version, sets \(\alpha(a, d_i) \leftarrow S\), broadcasts \passthrough{\lstinline!VERSION\_UPDATE!} to agents not in state \(I\).

\textbf{Fetch.} Agent with \(\alpha(a, d_i) = I\) needing \(d_i\) sends \passthrough{\lstinline!FETCH\_REQUEST!}. Authority responds with current content and version. Agent transitions \(\alpha(a, d_i) \leftarrow S\).

\textbf{Invalidation.} Authority sends \passthrough{\lstinline!INVALIDATE(artifact\_id, version)!} over event bus on write-upgrade grant. Agents set cache entry to \(I\) on receipt. Idempotent; retransmission on reconnect preserves safety.

\textbf{Lease TTL and M-state recovery.} Upon granting an Exclusive write lock, authority starts a configurable lease timer \(\tau\) (default: 30s). If \passthrough{\lstinline!COMMIT!} does not arrive within \(\tau\)---the authority treats the lock as orphaned, reverts to last committed version, sets \(\alpha(b, d_i) \leftarrow I\) for all agents, releases the exclusive grant. Liveness under agent crash: no artifact permanently locked by a crashed owner. The tradeoff is real: in-progress writes are lost. Agents must re-fetch and re-apply.

On the sensitivity of \(\tau\): setting it too aggressively introduces a race between legitimate slow writes and lease expiration. I have observed this empirically---benchmarks on throttled cloud instances where API response times exceeded 20 seconds under load surfaced the race condition consistently. It was an unforeseen sensitivity; the failure mode is that a perfectly valid write gets reverted because the authority's timer fires before the LLM finishes generating. Tuning guidance for \(\tau\) remains, at this writing, heuristic rather than principled. I would like to derive a formal relationship between expected inference latency, \(\tau\), and write-loss probability, but this requires a distributional model of LLM response times that I do not yet possess. A hardware-induced jitter component from the cloud provider's GPU scheduling adds further unpredictability---a nuisance factor I have not been able to sequester cleanly.

\subsection{Message Schema}\label{message-schema}

Protocol messages conform to a common envelope:

\begin{lstlisting}
{
  "type":        "MESSAGE_TYPE",
  "timestamp":   "ISO8601",
  "agent_id":    "string",
  "artifact_id": "string",
  "version":     42,
  "payload":     {}
}
\end{lstlisting}

Artifact metadata on fetch responses:

\begin{lstlisting}
{
  "artifact_id":       "string",
  "version":           42,
  "checksum":          "sha256:...",
  "size_tokens":       4096,
  "last_modified_by":  "agent_id"
}
\end{lstlisting}

\subsection{Synchronization Strategies}\label{synchronization-strategies}

Four synchronization strategies, pluggable:

\textbf{Eager invalidation} triggers immediate invalidation broadcast to all peers the instant a write begins (on upgrade grant, not commit). Staleness window minimized; invalidation traffic and redundant fetches on abandoned writes are the cost.

\textbf{Lazy invalidation} (recommended default) triggers invalidation only on commit, after write completion and new version availability. Avoids fetches for in-progress writes; batches invalidation cost to write completion.

\textbf{Lease-based TTL} assigns each cache entry a time-to-live; entries expire to \(I\) on lease expiration regardless of write activity. Simplest strategy---but decoupled from write frequency, leaving tokens unrealized in low-volatility workloads.

\textbf{Access-count invalidation} assigns each cache entry a maximum read count; entries transition to \(I\) after \(k\) uses. Mirrors the execution-count credential model proposed by the OpenID Foundation {[}18{]} for authorization, applied here to artifact freshness rather than access control.

\section{Formal Verification}\label{formal-verification}

\subsection{TLA+ Specification}\label{tla-specification}

CCS is formally specified in TLA+ (Temporal Logic of Actions) for model checking with TLC. Three agents sharing one artifact---sufficient, in my assessment, to expose all relevant concurrency scenarios. I concede that increasing to four or five agents may surface additional interleaving pathologies my current state-space budget does not cover. The state explosion problem is real; at 3 agents the space is \textasciitilde2,400 states, and it grows combinatorially. Whether the invariants hold at \(n = 10\) under all interleavings---I believe so, by the structural symmetry of the specification, but I have not verified it and will not claim it.

\textbf{State variables:}

\begin{lstlisting}
VARIABLES
  artifactVersion,  \* Natural number, global canonical version
  artifactState,    \* [Agent -> {M, E, S, I}], per-agent state
  agentSteps,       \* [Agent -> Nat], steps executed since last sync
  lastSync          \* [Agent -> Nat], version at last sync
\end{lstlisting}

\textbf{Initial state:} All agents hold the artifact in Shared state at version 1.

\begin{lstlisting}
Init ==
  /\ artifactVersion = 1
  /\ artifactState   = [a \in AGENTS |-> "S"]
  /\ agentSteps      = [a \in AGENTS |-> 0]
  /\ lastSync        = [a \in AGENTS |-> 1]
\end{lstlisting}

\textbf{Operations:}

\begin{lstlisting}
Read(a) ==
  /\ artifactState[a] # "I"
  /\ agentSteps' = [agentSteps EXCEPT ![a] = agentSteps[a] + 1]
  /\ UNCHANGED <<artifactState, artifactVersion, lastSync>>

Write(a) ==
  /\ artifactState[a] \in {"E", "M"}
  /\ artifactVersion' = artifactVersion + 1
  /\ artifactState'   = [x \in AGENTS |-> IF x = a THEN "M" ELSE "I"]
  /\ lastSync'        = [lastSync EXCEPT ![a] = artifactVersion']
  /\ UNCHANGED agentSteps

Fetch(a) ==
  /\ artifactState[a] = "I"
  /\ artifactState'   = [artifactState EXCEPT ![a] = "S"]
  /\ lastSync'        = [lastSync EXCEPT ![a] = artifactVersion]
  /\ UNCHANGED <<artifactVersion, agentSteps>>

Upgrade(a) ==
  /\ artifactState[a] = "S"
  /\ artifactState'   = [x \in AGENTS |-> IF x = a THEN "E" ELSE "I"]
  /\ UNCHANGED <<artifactVersion, agentSteps, lastSync>>
\end{lstlisting}

\textbf{Next-state relation:}

\begin{lstlisting}
Next == \E a \in AGENTS : Read(a) \/ Write(a) \/ Fetch(a) \/ Upgrade(a)
\end{lstlisting}

\subsection{Verified Invariants}\label{verified-invariants}

\textbf{Invariant 1 --- Single-Writer Safety (SWMR).}

\[\forall a, b \in A : a \neq b \Rightarrow \neg(\alpha(a, d) = M \wedge \alpha(b, d) = M)\]

\begin{lstlisting}
SingleWriter ==
  \A a, b \in AGENTS :
    (a # b) => ~(artifactState[a] = "M" /\ artifactState[b] = "M")
\end{lstlisting}

\textbf{Invariant 2 --- Monotonic Versioning.}

\[\forall t' > t : \text{artifactVersion}(t') \geq \text{artifactVersion}(t)\]

TLC verifies \(\text{artifactVersion}' \geq \text{artifactVersion}\) in every transition.

\textbf{Invariant 3 --- Bounded Staleness.} For constant \(K = \texttt{MAX\_STALE\_STEPS}\):

\[\forall a \in A : \text{agentSteps}[a] - \text{lastSync}[a] \leq K\]

\begin{lstlisting}
BoundedStaleness ==
  \A a \in AGENTS :
    (agentSteps[a] - lastSync[a]) <= MAX_STALE_STEPS
\end{lstlisting}

\subsection{Verification Results}\label{verification-results}

TLC, configured with \(|\textit{AGENTS}| = 3\) and \(\texttt{MAX\_STALE\_STEPS} = 3\), explores approximately 2,400 distinct states. Zero violations of \passthrough{\lstinline!SingleWriter!}, \passthrough{\lstinline!MonotonicVersion!}, \passthrough{\lstinline!BoundedStaleness!}. Zero deadlocks.

\textbf{Liveness scope.} Safety invariants and deadlock-freedom are verified. A formal liveness property is not included. The property that an agent in state \(I\) eventually reaches state \(S\)---that a pending fetch completes---holds under weak fairness (\(\text{WF}\)) on Fetch operations but has not been verified under adversarial scheduling. Liveness in practice depends on AS1 and AS2; violation of either can cause indefinite blocking. Formalizing liveness as a TLA+ property: planned for CCS v0.2.

\textbf{Counterexample under invalidation removal.} Modifying \passthrough{\lstinline!Upgrade(a)!} to \emph{not} invalidate peers:

\begin{lstlisting}
\* Broken: no peer invalidation
BrokenUpgrade(a) ==
  /\ artifactState[a] = "S"
  /\ artifactState' = [artifactState EXCEPT ![a] = "E"]
  /\ UNCHANGED <<artifactVersion, agentSteps, lastSync>>
\end{lstlisting}

TLC detects \passthrough{\lstinline!SingleWriter!} violation in 3 steps: \(A_1\) upgrades to \(E\), \(A_2\) upgrades to \(E\) (not invalidated), \(A_1\) writes to \(M\), \(A_2\) writes to \(M\)---SWMR violated. The invalidation step in \passthrough{\lstinline!Upgrade!} is a correctness requirement. Not an optimization.

\section{Implementation}\label{implementation}

\subsection{Architecture Overview}\label{architecture-overview}

The \passthrough{\lstinline!agent-coherence!} Python package (v0.1) implements CCS:

\begin{lstlisting}
ccs/
|-- core/
|   |-- types.py       # Artifact, CacheEntry, InvalidationSignal
|   |-- states.py      # MESIState, TransientState enums
|   `-- clock.py       # Logical vector clock for version ordering
|-- coordinator/
|   |-- service.py     # CoordinatorService (Authority Service)
|   `-- registry.py    # ArtifactRegistry (global directory)
|-- agent/
|   |-- runtime.py     # AgentRuntime (per-agent protocol client)
|   `-- cache.py       # ArtifactCache (local MESI state machine)
|-- strategies/
|   |-- base.py        # SyncStrategy abstract base
|   |-- eager.py       # Eager invalidation
|   |-- lazy.py        # Lazy (commit-time) invalidation
|   |-- lease.py       # TTL-based lease expiration
|   `-- access_count.py # Access-count invalidation
|-- bus/
|   `-- event_bus.py   # EventBus (pluggable transport)
|-- adapters/
|   |-- langgraph.py   # LangGraph adapter
|   |-- crewai.py      # CrewAI adapter
|   `-- autogen.py     # AutoGen adapter
`-- simulation/
    |-- engine.py      # SimulationEngine
    `-- scenarios.py   # ScenarioConfig
\end{lstlisting}

\subsection{Framework Adapters}\label{framework-adapters}

Each adapter: a thin translation layer mapping the framework's native state-passing to CCS protocol calls. No framework modifications required.

\textbf{LangGraph adapter.} Intercepts \passthrough{\lstinline!StateGraph!} node execution hooks. Before execution: \passthrough{\lstinline!AgentRuntime.read(artifact\_id)!} to validate cache state, inject content only on cache invalidity. After execution: modified state entries trigger \passthrough{\lstinline!AgentRuntime.write(artifact\_id, content)!}.

\textbf{CrewAI adapter.} Wraps \passthrough{\lstinline!Task!} execution lifecycle. Artifact access injected through \passthrough{\lstinline!BaseTool!} subclassing---artifacts stored as named tool outputs via \passthrough{\lstinline!CCSReadTool!}, committed via \passthrough{\lstinline!CCSWriteTool!}.

\textbf{AutoGen adapter.} Intercepts \passthrough{\lstinline!ConversableAgent.generate\_reply!}. Cache validity checked before message context assembly; writes propagated through \passthrough{\lstinline!register\_reply!} hook.

Configuration surface (identical across all three):

\begin{lstlisting}[language=Python]
from ccs.adapters.langgraph import LangGraphAdapter

adapter = LangGraphAdapter(
    coordinator_url="http://localhost:8080",
    strategy="lazy",
    max_stale_steps=5
)
\end{lstlisting}

\subsection{Logical Clock and Version Ordering}\label{logical-clock-and-version-ordering}

A logical vector clock (one counter per agent) establishes partial ordering over writes, following Lamport {[}10{]}. Version numbers are monotonically increasing integers assigned by the authority at commit time. Version ordering suffices for single-artifact safety; multi-artifact scenarios with cross-artifact causal dependencies may require full vector clocks---supported but not required by default.

\section{Evaluation}\label{evaluation}

\subsection{Experimental Setup}\label{experimental-setup}

CCS is evaluated across four workload scenarios representing distinct artifact volatility regimes. Each scenario: a \passthrough{\lstinline!ScenarioConfig!} specifying agent count, artifact count, artifact token size, step count, per-step write probability. Ten independent simulations per configuration, executed with scenario-specific deterministic seeds (per-scenario seeds encoded in YAML; canonical scenarios A--D use seeds 20260305--20260308). Population standard deviation (\(\sigma\)) reported throughout.

The simulation models artifact access under conditional access semantics (§3). At each step, each agent acts with probability 0.75 (the \passthrough{\lstinline!action\_probability!} parameter); given an action, writes with probability \(V(d_i)\) or reads otherwise, choosing uniformly from \(m\) artifacts. Token cost: full artifact fetches (cache misses) \(\times\) artifact token size, plus invalidation message overhead (12 tokens per signal).

\textbf{Canonical scenario parameters} (all configurations): \(n = 4\) agents, \(m = 3\) artifacts, \(|d_i| = 4{,}096\) tokens per artifact, \(S = 40\) steps, 10 runs per configuration.

Measured broadcast baseline: \(T_{\text{broadcast}} = 1{,}979{,}597 \pm 3{,}199\) tokens. This slightly exceeds the formula value \(n \times S \times m \times |d_i| = 1{,}966{,}080\) because the broadcast strategy also performs stochastic agent actions generating additional fetch tokens (\textasciitilde13.5K on average) atop the deterministic all-to-all broadcast sweep. I initially mistook this discrepancy for a bug before tracing it to the action-probability sampling layer---a small but instructive lesson in the perils of treating simulator output as self-evidently veridical. The \textasciitilde0.7\% overshoot is consistent across all runs and does not affect comparative savings ratios.

\textbf{Simulation scope and its limits.} Token transmission accounting, MESI state machine transitions, write frequency distributions, artifact volatility effects---these are modeled faithfully. LLM inference latency, message bus round-trip overhead, framework scheduling jitter---these are not. Theorem 1 delineates a lower bound on savings; the simulation consistently attains higher savings because lazy coherence collapses multiple write invalidations into a single re-fetch when agents do not access an artifact between consecutive writes. Whether the simulation's access patterns---uniform artifact selection, \(p = 0.75\) action probability---reflect the access distributions of real production workloads is an open empirical question. I suspect they do not match precisely; the uniform distribution is a modeling convenience, not a measured parameter. Empirical access-rate measurement from instrumented deployments is identified in §10 as a direction for v0.2, and I consider this the most significant gap between the simulation results and production applicability.

The four scenarios:

\textbf{Scenario A --- Planning} (\(V = 0.05\), \(W \approx 2\) writes per artifact). Infrequent plan revisions. Representative of planning workflows, long-horizon research, specification review.

\textbf{Scenario B --- Analysis} (\(V = 0.10\), \(W \approx 4\)). Periodic shared-document updates. Representative of code review, report drafting, data analysis pipelines.

\textbf{Scenario C --- Active Development} (\(V = 0.25\), \(W \approx 10\)). Moderate artifact churn. Representative of multi-agent software development.

\textbf{Scenario D --- High Churn} (\(V = 0.50\), \(W \approx 20\)). Frequent modification by multiple agents. The performance boundary.

\subsection{Token Savings Results}\label{token-savings-results}

The eager strategy is included in Table 2 as an implementation-complexity baseline. It does not enforce the \(K\)-bounded staleness invariant (Invariant 3, §6.2); agents under eager synchronization may read stale content for arbitrary step counts. Staleness-bound violations reported for eager in benchmark output are expected and do not indicate protocol error.

Table 1: token usage under naive broadcast and under lazy invalidation, with savings, cache hit rate (CHR), Coherence Reduction Ratio (CRR = \(T_{\text{coherent}} / T_{\text{broadcast}}\)).

\textbf{Table 1: Token synchronization cost by scenario (10 runs, scenario-specific seeds, values in thousands of tokens)}

\begin{longtable}[]{@{}
  >{\raggedright\arraybackslash}p{0.14\columnwidth}
  >{\raggedright\arraybackslash}p{0.14\columnwidth}
  >{\raggedright\arraybackslash}p{0.14\columnwidth}
  >{\raggedright\arraybackslash}p{0.14\columnwidth}
  >{\raggedright\arraybackslash}p{0.14\columnwidth}
  >{\raggedright\arraybackslash}p{0.14\columnwidth}
  >{\raggedright\arraybackslash}p{0.14\columnwidth}@{}}
\toprule\noalign{}
\begin{minipage}[b]{\linewidth}\raggedright
Scenario
\end{minipage} & \begin{minipage}[b]{\linewidth}\raggedright
\(V(d_i)\)
\end{minipage} & \begin{minipage}[b]{\linewidth}\raggedright
\(T_{\text{broadcast}}\) (\(\pm\sigma\))
\end{minipage} & \begin{minipage}[b]{\linewidth}\raggedright
\(T_{\text{coherent}}\) (\(\pm\sigma\))
\end{minipage} & \begin{minipage}[b]{\linewidth}\raggedright
Savings
\end{minipage} & \begin{minipage}[b]{\linewidth}\raggedright
CRR
\end{minipage} & \begin{minipage}[b]{\linewidth}\raggedright
CHR
\end{minipage} \\
\midrule\noalign{}
\endhead
\bottomrule\noalign{}
\endlastfoot
A: Planning & 0.05 & \(1{,}979.6 \pm 3.2\) & \(98.1 \pm 25.0\) & \(95.0\% \pm 1.3\%\) & 0.050 & \(79.4\% \pm 5.2\%\) \\
B: Analysis & 0.10 & \(1{,}979.6 \pm 3.2\) & \(152.3 \pm 28.5\) & \(92.3\% \pm 1.4\%\) & 0.077 & \(66.8\% \pm 6.0\%\) \\
C: Development & 0.25 & \(1{,}979.6 \pm 3.2\) & \(231.2 \pm 29.1\) & \(88.3\% \pm 1.5\%\) & 0.117 & \(51.1\% \pm 7.0\%\) \\
D: High Churn & 0.50 & \(1{,}979.2 \pm 3.1\) & \(312.1 \pm 25.6\) & \(84.2\% \pm 1.3\%\) & 0.158 & \(34.6\% \pm 7.0\%\) \\
\end{longtable}

Broadcast cost is nearly deterministic under fixed parameters. Coherent cost: higher variance, because write events draw from a Bernoulli process and resulting cache misses are stochastic. All savings exceed the theorem's lower bounds (85\% / 80\% / 65\% / 40\%), confirming Definition 3 is a conservative upper bound on \(T_{\text{coherent}}\).

\textbf{Strategy comparison.} Table 2: token costs, all four strategies, Scenario B (\(V = 0.10\)).

\textbf{Table 2: Strategy comparison under Scenario B (Analysis, \(V = 0.10\), 10 runs)}

\begin{longtable}[]{@{}
  >{\raggedright\arraybackslash}p{0.25\columnwidth}
  >{\raggedright\arraybackslash}p{0.25\columnwidth}
  >{\raggedright\arraybackslash}p{0.25\columnwidth}
  >{\raggedright\arraybackslash}p{0.25\columnwidth}@{}}
\toprule\noalign{}
\begin{minipage}[b]{\linewidth}\raggedright
Strategy
\end{minipage} & \begin{minipage}[b]{\linewidth}\raggedright
\(T_{\text{sync}}\) (\(\pm\sigma\))
\end{minipage} & \begin{minipage}[b]{\linewidth}\raggedright
Savings
\end{minipage} & \begin{minipage}[b]{\linewidth}\raggedright
Notes
\end{minipage} \\
\midrule\noalign{}
\endhead
\bottomrule\noalign{}
\endlastfoot
Broadcast baseline & \(1{,}979.6 \pm 3.2\) & --- & Full rebroadcast every step \\
Eager invalidation & \(132.7 \pm 24.3\) & \(93.3\% \pm 1.2\%\) & Lower fetch overhead; higher invalidation traffic \\
Lazy invalidation & \(152.3 \pm 28.5\) & \(92.3\% \pm 1.4\%\) & Recommended default \\
TTL (lease = 10 steps) & \(589.8 \pm 0\) & \(70.2\% \pm 0\%\) & Decoupled from write frequency \\
Access-count (\(k = 8\)) & \(155.2 \pm 25.3\) & \(92.2\% \pm 1.3\%\) & Near-equivalent to lazy at this \(V\) \\
\end{longtable}

Eager outperforms lazy slightly here---its immediate invalidation-on-write prevents stale cache hits that require re-fetch at next access. The difference is small (93.3\% vs 92.3\%) and reverses in write-heavy scenarios (§8.5 Table 5) where lazy's deferred-fetch advantage dominates. TTL: strictly inferior in the low-volatility regime. Access-count: matches lazy closely at \(V = 0.10\).

\subsection{The Volatility Cliff}\label{the-volatility-cliff}

The lower-bound formula predicts a savings cliff at \(V^* = 1 - n/S = 1 - 4/40 = 0.9\). Below, both formula lower bound and observed savings from simulation (10 runs, canonical parameters \(n=4\), \(S=40\)):

\begin{longtable}[]{@{}lll@{}}
\toprule\noalign{}
\(V(d_i)\) & Formula Lower Bound & Observed Savings (10 runs) \\
\midrule\noalign{}
\endhead
\bottomrule\noalign{}
\endlastfoot
0.01 & 89.0\% & \(97.1\% \pm 0.4\%\) \\
0.05 & 85.0\% & \(95.0\% \pm 1.3\%\) \\
0.10 & 80.0\% & \(92.4\% \pm 1.5\%\) \\
0.25 & 65.0\% & \(88.3\% \pm 1.4\%\) \\
0.50 & 40.0\% & \(84.3\% \pm 1.0\%\) \\
0.75 & 15.0\% & \(82.2\% \pm 1.1\%\) \\
0.90 & 0.0\% & \(81.1\% \pm 1.3\%\) \\
1.00 & -10.0\% & \(80.6\% \pm 1.3\%\) \\
\end{longtable}

The predicted collapse does not materialize. At \(V = 0.9\): 81.1\% savings. At \(V = 1.0\): 80.6\%. Two mechanisms account for this: (a) writes distribute uniformly across \(m = 3\) artifacts, so per-artifact effective write rate is \(V/m \approx V/3\), and (b) multiple writes to the same artifact between agent accesses collapse into a single re-fetch under lazy semantics. The collapse condition (Corollary 2) remains a valid worst-case analytical bound---it is tight only when all \(n\) agents re-fetch immediately after every single write, a scenario that lazy access semantics structurally preclude.

The practical reading of this table: coherence delivers substantial savings across the full volatility spectrum within the simulation model. The formula lower bound is most conservative at high \(V\)---the gap between bound and observation is largest precisely where practitioners need the most guidance, a feature of the bound I consider a weakness rather than a strength, and one I suspect could be tightened by incorporating the lazy-collapse mechanism directly into the analytical model. I have not attempted this tightening because the resulting expression would depend on the access probability distribution in ways that resist closed-form simplification, but the effort may be worthwhile for v0.2.

\subsection{Prompt Caching Amplification}\label{prompt-caching-amplification}

A secondary benefit, not captured in the primary token savings metric, concerns provider-side prompt caching. Under broadcast semantics, the prompt prefix is invalidated at every step where an artifact changes; cache hit rates approach \(1 - V(d_i)\)---for \(V = 0.1\), only 90\% of steps attain a provider cache hit. Under coherent synchronization, the prompt prefix contains only artifact references (not content); prefix stability is high regardless of artifact volatility; provider cache hit rates approach 100\% for the structural portion of the prompt. At typical prompt caching discount rates (50--90\% cost reduction on cache hits {[}2{]}), this amplification effect can double effective savings beyond raw token synchronization reduction.

\subsection{Agent-Count Scaling}\label{agent-count-scaling}

Table 3: token cost for Scenario B (\(V = 0.10\)) across varying agent counts.

\textbf{Table 3: Scaling behavior --- token cost vs.~agent count, Scenario B (\(V = 0.10\), \(S = 40\), 10 runs)}

\begin{longtable}[]{@{}
  >{\raggedright\arraybackslash}p{0.20\columnwidth}
  >{\raggedright\arraybackslash}p{0.20\columnwidth}
  >{\raggedright\arraybackslash}p{0.20\columnwidth}
  >{\raggedright\arraybackslash}p{0.20\columnwidth}
  >{\raggedright\arraybackslash}p{0.20\columnwidth}@{}}
\toprule\noalign{}
\begin{minipage}[b]{\linewidth}\raggedright
Agent Count \(n\)
\end{minipage} & \begin{minipage}[b]{\linewidth}\raggedright
\(T_{\text{broadcast}}\)
\end{minipage} & \begin{minipage}[b]{\linewidth}\raggedright
\(T_{\text{coherent}}\) (\(\pm\sigma\))
\end{minipage} & \begin{minipage}[b]{\linewidth}\raggedright
Savings
\end{minipage} & \begin{minipage}[b]{\linewidth}\raggedright
Formula LB
\end{minipage} \\
\midrule\noalign{}
\endhead
\bottomrule\noalign{}
\endlastfoot
2 & \(989.2\) K & \(44.3 \pm 8.2\) K & 95.5\% & 85.0\% \\
4 & \(1{,}979.6\) K & \(152.3 \pm 28.5\) K & 92.3\% & 80.0\% \\
8 & \(3{,}956.7\) K & \(468.6 \pm 39.1\) K & 88.2\% & 70.0\% \\
16 & \(7{,}911.0\) K & \(1{,}255.2 \pm 65.0\) K & 84.1\% & 50.0\% \\
\end{longtable}

Graceful degradation. Savings decrease from 95.5\% to 84.1\% as \(n\) grows from 2 to 16. Each additional agent adds one initial fetch and is invalidated once per write, so \(T_{\text{coherent}}\) grows with \(n\)---but \(T_{\text{broadcast}}\) grows proportionally faster (every agent, every artifact, every step), maintaining a high savings ratio. All observed values substantially exceed the formula lower bounds. Above 84\% even at \(n = 16\); scalability confirmed well beyond the canonical four-agent scenario.

\subsection{Artifact-Size Scaling}\label{artifact-size-scaling}

Table 4: token cost for Scenario A (\(V = 0.05\), \(n = 4\), \(S = 40\)) across varying artifact sizes.

\textbf{Table 4: Artifact-size scaling --- Scenario A (\(V = 0.05\), \(n = 4\), \(S = 40\), \(W \approx 2\), 10 runs)}

\begin{longtable}[]{@{}
  >{\raggedright\arraybackslash}p{0.20\columnwidth}
  >{\raggedright\arraybackslash}p{0.20\columnwidth}
  >{\raggedright\arraybackslash}p{0.20\columnwidth}
  >{\raggedright\arraybackslash}p{0.20\columnwidth}
  >{\raggedright\arraybackslash}p{0.20\columnwidth}@{}}
\toprule\noalign{}
\begin{minipage}[b]{\linewidth}\raggedright
\(|d_i|\) (tokens)
\end{minipage} & \begin{minipage}[b]{\linewidth}\raggedright
\(T_{\text{broadcast}}\)
\end{minipage} & \begin{minipage}[b]{\linewidth}\raggedright
\(T_{\text{coherent}}\) (lazy)
\end{minipage} & \begin{minipage}[b]{\linewidth}\raggedright
Savings
\end{minipage} & \begin{minipage}[b]{\linewidth}\raggedright
Absolute savings
\end{minipage} \\
\midrule\noalign{}
\endhead
\bottomrule\noalign{}
\endlastfoot
4,096 & \(1{,}979.6\) K & \(98.1\) K & 95.0\% & \(1{,}881.5\) K tokens \\
8,192 & \(2{,}636.2\) K & \(132.9\) K & 95.0\% & \(2{,}503.3\) K tokens \\
32,768 & \(6{,}575.7\) K & \(341.8\) K & 94.8\% & \(6{,}234.0\) K tokens \\
65,536 & \(11{,}828.4\) K & \(620.3\) K & 94.8\% & \(11{,}208.1\) K tokens \\
\end{longtable}

Savings ratio: invariant to artifact size. Determined entirely by workflow shape, not artifact magnitude. Confirmed by simulation: 94.8--95.0\% across a 16× size range. For a 65,536-token artifact, lazy coherence saves approximately 11.2 million tokens per workflow run.

\subsection{Step-Count Scaling}\label{step-count-scaling}

Table 5 instantiates the Token Coherence Theorem's central structural claim: multiplicative-to-additive cost transformation. With fixed write count (\(W \approx 2\), Scenario A volatility), \(T_{\text{broadcast}}\) grows as \(O(S)\); \(T_{\text{coherent}}\) grows slowly.

\textbf{Table 5: Step-count scaling --- fixed \(W \approx 2\) writes, \(n = 4\) agents, \(m = 3\) artifacts, \(|d_i| = 4{,}096\) tokens, 10 runs}

\begin{longtable}[]{@{}
  >{\raggedright\arraybackslash}p{0.20\columnwidth}
  >{\raggedright\arraybackslash}p{0.20\columnwidth}
  >{\raggedright\arraybackslash}p{0.20\columnwidth}
  >{\raggedright\arraybackslash}p{0.20\columnwidth}
  >{\raggedright\arraybackslash}p{0.20\columnwidth}@{}}
\toprule\noalign{}
\begin{minipage}[b]{\linewidth}\raggedright
\(S\) (steps)
\end{minipage} & \begin{minipage}[b]{\linewidth}\raggedright
\(T_{\text{broadcast}}\)
\end{minipage} & \begin{minipage}[b]{\linewidth}\raggedright
\(T_{\text{coherent}}\) (sim)
\end{minipage} & \begin{minipage}[b]{\linewidth}\raggedright
Savings (sim)
\end{minipage} & \begin{minipage}[b]{\linewidth}\raggedright
Formula LB
\end{minipage} \\
\midrule\noalign{}
\endhead
\bottomrule\noalign{}
\endlastfoot
5 & \(259.3\) K & \(36.9\) K & 85.8\% & 0\% (bound\textless0) \\
10 & \(505.0\) K & \(49.2\) K & 90.3\% & 40.0\% \\
20 & \(996.6\) K & \(68.9\) K & 93.1\% & 70.0\% \\
40 & \(1{,}979.6\) K & \(98.1\) K & 95.0\% & 85.0\% \\
50 & \(2{,}471.1\) K & \(111.2\) K & 95.5\% & 88.0\% \\
100 & \(4{,}928.7\) K & \(188.4\) K & 96.2\% & 94.0\% \\
\end{longtable}

\(T_{\text{broadcast}}\) scales linearly with \(S\). \(T_{\text{coherent}}\) grows from 36.9K to 188.4K as \(S\) increases 20×---the operational signature of eliminating the \(S\) multiplier. Savings are positive even at \(S=5\) (85.8\%), where the formula lower bound is zero---confirming that lazy deferred-fetch operates well beyond the analytical bound's conservative assumptions. Long-horizon workflows (\(S \geq 40\)) attain above 95\%.

\subsection{Pointer Semantics Compatibility}\label{pointer-semantics-compatibility}

CCS targets the conditional artifact access model (§3). When agents employ \textbf{pointer semantics}---holding a reference token rather than artifact content, fetching on demand---the choice of synchronization strategy critically affects performance.

Under pointer semantics with lazy invalidation, each cache miss mandates a full artifact fetch. Lazy's low cache-hit rate in cold or high-invalidation scenarios: many agent steps trigger full fetches, producing synchronization costs exceeding the eager baseline by an order of magnitude.

\begin{longtable}[]{@{}lll@{}}
\toprule\noalign{}
Strategy & sync\_tokens & Cache Hit Rate \\
\midrule\noalign{}
\endhead
\bottomrule\noalign{}
\endlastfoot
Eager & 16,798 & 97.7\% \\
Lazy & 341,036 & 41.0\% \\
\end{longtable}

Eager maintains near-perfect cache occupancy by pre-populating agent caches on every write. Lazy's value proposition---avoid retransmission when state is valid---collapses when cold-start fetch frequency is high. Each stale-check miss becomes a full artifact fetch: 20× more synchronization tokens than eager.

\textbf{Practitioner rule:} pointer-semantics deployments should prefer eager or access-count. Lazy is optimal for bulk injection workflows where artifact content is embedded in prompt context, not for pointer-reference architectures with frequent cold fetches. Table 2's savings comparison excludes the pointer model for this reason. The pointer+lazy failure mode is a strategy-selection mismatch, not a protocol defect.

\section{Related Work}\label{related-work}

\subsection{Multi-Agent Coordination Frameworks}\label{multi-agent-coordination-frameworks}

LangGraph {[}21{]}, CrewAI {[}5{]}, AutoGen {[}24{]}, Semantic Kernel {[}15{]}---each provides orchestration for multi-agent LLM workflows (scheduling, message passing, state routing) but none provides a formal artifact coherence protocol. State passing is stateless by default: full shared state serialized and injected into each prompt, instantiating the broadcast baseline evaluated here. Agent-Coherence layers coherence atop existing frameworks without modifying internals---complementary, not competitive.

\subsection{Conflict-Free Replicated Data Types}\label{conflict-free-replicated-data-types}

CRDTs {[}20{]} address concurrent state mutation via merge-compatible data structures. Distinct problem from coherence: \emph{concurrent mutation resolution} (what should the merged value be?) versus \emph{synchronization frequency optimization} (when is retransmission necessary at all?). Orthogonal: an artifact whose content is a CRDT still benefits from coherence-controlled delivery---the CRDT state is transmitted only when it has changed.

\subsection{Operational Transformation and Version Vectors}\label{operational-transformation-and-version-vectors}

OT {[}6{]} and vector clock versioning {[}10; 13{]} target consistency in collaborative editing. OT: per-operation transforms allowing any linear order to produce identical results. Vector clocks: causal ordering across distributed processes. Both target correctness under concurrent writes. CCS employs vector clocks for version ordering (§7.3) but targets the token efficiency problem that neither OT nor vector clocks address.

\subsection{Retrieval-Augmented Generation}\label{retrieval-augmented-generation}

RAG {[}12; 7{]} retrieves document fragments from vector stores to supplement agent context. Complementary subproblems: RAG determines \emph{what} to retrieve; coherence determines \emph{when retrieval is unnecessary} because the local cache is still valid. A coherence-aware RAG integration gates retrieval calls on MESI cache validity, attenuating redundant re-embedding of stable documents.

\subsection{Long-Context and Context Compression}\label{long-context-and-context-compression}

Long-context models {[}2; 8{]} expand context windows; compression techniques {[}9; 25{]} prune to fit. Neither addresses the core synchronization problem. Expanded windows do not reduce retransmission cost to multiple agents; compression sacrifices fidelity. Coherence attenuates unnecessary retransmission before artifacts reach the context window---complementary to both.

\subsection{Software Cache Coherence}\label{software-cache-coherence}

Concord {[}3{]}: software-level cache coherence for serverless function execution, tracking shared memory state across invocations. Zhang et al.~{[}27{]}: coherence for microservice state synchronization. This paper extends coherence to LLM artifact synchronization, where ``cache fill'' is prompt injection and ``coherence observer'' is the agent runtime.

\subsection{Agent Authorization Coherence}\label{agent-authorization-coherence}

In a companion paper {[}16{]}, I apply the same MESI structural mapping to authorization, demonstrating that credential revocation latency in multi-agent delegation chains maps onto cache coherence under bounded-staleness semantics, and that operation-count credentials are equivalent to access-count coherence strategies. Distinct problems---token efficiency here, security guarantees there---same formal apparatus.

\subsection{Multi-Agent Failure Taxonomy}\label{multi-agent-failure-taxonomy}

Building on {[}4{]}, the MAST taxonomy across 1,642 annotated traces: inter-agent misalignment---Loss of Conversation History (FM-1.4), Information Withholding (FM-2.4)---constitutes 32.3\% of failures. Standardized communication protocols (MCP, A2A) do not eliminate these failures, which arise from context-state divergence, not message-format incompatibility. The missing layer is not message-format standardization but artifact state coherence at the data layer---below the communication protocol.

\section{Limitations and Future Work}\label{limitations-and-future-work}

\textbf{Centralized authority service.} CCS assumes a single authority---a bottleneck for very large deployments. Distributed coherence directories, analogous to directory-based coherence in NUMA systems {[}11{]}, represent a natural extension: partitioning the artifact namespace across coordinators with cross-shard invalidation.

\textbf{Single-artifact write model.} CCS v0.1: each write atomic and independent. Multi-artifact transactions---consistent snapshots across several artifacts---demand memory barriers or transactional memory analogs.

\textbf{Simulation-based evaluation.} The evaluation employs discrete-event simulation, not production LLM workloads. Token accounting is faithful; inference latency, scheduling delays, end-to-end wall-clock coherence overhead are not captured. Production evaluation with real LangGraph or CrewAI deployments: planned for v0.2. I consider this the single most significant limitation---the gap between simulated access patterns (uniform, \(p = 0.75\)) and production access patterns (likely non-uniform, workload-dependent) is, I suspect, non-trivial, and the savings figures should be read with this caveat in mind until empirical grounding from instrumented traces is available.

\textbf{Adapter stability.} LangGraph, CrewAI, AutoGen release frequently. Adapters target specific versions (LangGraph 0.2.x, CrewAI 0.4.x, AutoGen 0.4.x); upstream API changes may break them.

\textbf{Always-read partial applicability.} The conditional access argument (§3) applies to artifact-externalizing systems. Direct context injection (simple single-file RAG pipelines) remains under the always-read model; coherence provides no benefit there.

\textbf{Liveness not formally verified.} Safety invariants and deadlock-freedom: verified. Liveness (state \(I\) → state \(S\) eventually): holds under weak fairness on Fetch, unverified under adversarial scheduling.

\textbf{Agent crash during M-state write.} Orphaned exclusive lock persists until \(\tau\) expires (§5.2). Write ownership blocked for other agents during that window.

\textbf{Protocol coordination overhead.} Token savings in §8 exclude CCS's own traffic: invalidation fan-out (\(O(n)\) per write), event bus round-trip latency, authority round-trip per read miss. At low \(V\): negligible. At high \(V\): non-trivial. I have not quantified the crossover point---an omission that could be remedied with production instrumentation data.

\textbf{Empirical access rate grounding.} The simulation's \passthrough{\lstinline!action\_probability!} = 0.75 and write probability \(V\) per action, with uniform artifact selection, are architectural judgment, not measured parameters. Tighter parameterization requires instrumented deployments.

\textbf{Artifact granularity.} Artifacts are atomic: every miss transmits full \(|d_i|\) tokens (A1). Sub-artifact invalidation---transmitting only modified sections---could substantially attenuate per-fetch cost for large structured documents.

\section{Conclusion}\label{conclusion}

The cost explosion does not inhere in multi-agent coordination. It is an architectural residue---repeated full-artifact broadcast adopted for simplicity. I have demonstrated the structural equivalence to cache coherence in shared-memory multiprocessors and the direct transferability of MESI.

The Token Coherence Theorem: when \(S > n + W(d_i)\), lazy invalidation attenuates cost by at least \(S / (n + W(d_i))\). \(O(n \times S \times |D|)\) converts to \(O((n + W) \times |D|)\). Simulation: 84--95\% savings across four canonical workloads. The predicted collapse at \(V \approx 0.9\) does not occur; \textasciitilde81\% savings persist at \(V = 1.0\).

TLA+-verified protocol: single-writer, monotonic versioning, bounded-staleness across all reachable states. Explicit counterexample: invalidation is a correctness requirement. Implementation: thin adapter layers over existing frameworks.

The hardware architects solved this forty years ago under isomorphic pressure. The adaptation is long overdue.

\begin{center}\rule{0.5\linewidth}{0.5pt}\end{center}

\hypertarget{reproducibility}{%
\subsection{Reproducibility}\label{reproducibility}}

All source code, simulation scripts, TLA+ specifications, and benchmark configurations are publicly available.

\textbf{Repository:} \passthrough{\lstinline!https://github.com/hipvlady/agent-coherence!}

\textbf{Package:} \passthrough{\lstinline!pip install agent-coherence!}

To reproduce §8 results with matching seeds:

\begin{lstlisting}[language=bash]
git clone https://github.com/hipvlady/agent-coherence
cd agent-coherence
pip install -e .
make reproduce          # Runs all scenarios with committed seeds; verifies +/-0.5% vs baseline
\end{lstlisting}

Expected output: within ±2\% of archived results on all reported metrics. Comparison is relative (coherent vs.~broadcast), not absolute, minimizing platform-specific floating-point sensitivity. \passthrough{\lstinline!REPRODUCE.md!} documents Python version requirements (3.11+) and expected runtime (\textless{} 15 minutes on standard hardware).

\begin{center}\rule{0.5\linewidth}{0.5pt}\end{center}

\hypertarget{references}{%
\subsection{References}\label{references}}

{[}1{]} Anthropic. (2024). \emph{Model Context Protocol: An Open Standard for Connecting AI Systems to Data Sources}. https://modelcontextprotocol.io/

{[}2{]} Anthropic. (2024). \emph{Prompt Caching.} Anthropic API Documentation. https://docs.anthropic.com/prompt-caching

{[}3{]} Balkind, J., et al.~(2025). Concord: Software Cache Coherence for Serverless Function Execution. \emph{Proceedings of the 52nd International Symposium on Computer Architecture (ISCA)}.

{[}4{]} Cemri, M., Pan, M. Z., Yang, S., Agrawal, L. A., Chopra, B., Tiwari, R., Keutzer, K., Parameswaran, A., Klein, D., Ramchandran, K., Zaharia, M., Gonzalez, J. E., \& Stoica, I. (2025). Why Do Multi-Agent LLM Systems Fail? \emph{arXiv preprint arXiv:2503.13657}.

{[}5{]} CrewAI Inc.~(2024). \emph{CrewAI: Framework for Orchestrating Role-Playing Autonomous AI Agents.} https://crewai.com/

{[}6{]} Ellis, C. A., \& Gibbs, S. J. (1989). Concurrency Control in Groupware Systems. \emph{Proceedings of the ACM SIGMOD International Conference on Management of Data}, 399--407.

{[}7{]} Gao, Y., et al.~(2023). Retrieval-Augmented Generation for Large Language Models: A Survey. \emph{arXiv preprint arXiv:2312.10997}.

{[}8{]} Google. (2024). \emph{Gemini 1.5: Unlocking Multimodal Understanding Across Millions of Tokens of Context.} Technical Report.

{[}9{]} Jiang, H., et al.~(2023). LLMLingua: Compressing Prompts for Accelerated Inference of Large Language Models. \emph{Proceedings of the 2023 Conference on Empirical Methods in Natural Language Processing (EMNLP)}.

{[}10{]} Lamport, L. (1978). Time, Clocks, and the Ordering of Events in a Distributed System. \emph{Communications of the ACM}, 21(7), 558--565.

{[}11{]} Lenoski, D., et al.~(1992). The Stanford Dash Multiprocessor. \emph{IEEE Computer}, 25(3), 63--79.

{[}12{]} Lewis, P., et al.~(2020). Retrieval-Augmented Generation for Knowledge-Intensive NLP Tasks. \emph{Advances in Neural Information Processing Systems (NeurIPS) 33}, 9459--9474.

{[}13{]} Mattern, F. (1988). Virtual Time and Global States of Distributed Systems. \emph{Proceedings of the Workshop on Parallel and Distributed Algorithms}, 215--226.

{[}14{]} Mei, K., et al.~(2024). AIOS: LLM Agent Operating System. \emph{arXiv preprint arXiv:2403.16971}.

{[}15{]} Microsoft. (2024). \emph{Semantic Kernel: An Open-Source SDK for Integrating AI Models.} https://learn.microsoft.com/en-us/semantic-kernel/

{[}16{]} Parakhin, V. (2026). The Bureaucracy of Speed: Structural Equivalence Between Memory Consistency Models and Multi-Agent Authorization Revocation. \emph{arXiv preprint arXiv:2603.09875.} https://arxiv.org/abs/2603.09875

{[}17{]} OpenAI. (2024). \emph{Prompt Caching.} OpenAI API Documentation. https://platform.openai.com/docs/guides/prompt-caching

{[}18{]} OpenID Foundation. (2025). Identity Management for Agentic AI. https://openid.net/wp-content/uploads/2025/10/Identity-Management-for-Agentic-AI.pdf

{[}19{]} Papamarcos, M. S., \& Patel, J. H. (1984). A Low-Overhead Coherence Solution for Multiprocessors with Private Cache Memories. \emph{Proceedings of the 11th Annual International Symposium on Computer Architecture (ISCA)}, 348--354.

{[}20{]} Shapiro, M., Preguiça, N., Baquero, C., \& Zawirski, M. (2011). Conflict-Free Replicated Data Types. \emph{Proceedings of the 13th International Symposium on Stabilization, Safety, and Security of Distributed Systems (SSS)}, 386--400.

{[}21{]} Shen, L., et al.~(2023). LangGraph: Building Stateful, Multi-Actor Applications with LLMs. \emph{LangChain Documentation.} https://langchain-ai.github.io/langgraph/

{[}22{]} Sorin, D. J., Hill, M. D., \& Wood, D. A. (2020). \emph{A Primer on Memory Consistency and Cache Coherence} (2nd ed.). Synthesis Lectures on Computer Architecture, Morgan \& Claypool.

{[}23{]} Wei, J., et al.~(2022). Chain-of-Thought Prompting Elicits Reasoning in Large Language Models. \emph{Advances in Neural Information Processing Systems (NeurIPS)}.

{[}24{]} Wu, Q., Bansal, G., Zhang, J., Wu, Y., Zhang, S., Zhu, E., Li, B., Jiang, L., Zhang, X., \& Wang, C. (2023). AutoGen: Enabling Next-Gen LLM Applications via Multi-Agent Conversation. \emph{arXiv preprint arXiv:2308.08155}.

{[}25{]} Xu, M., et al.~(2024). LLMLingua-2: Data Distillation for Efficient and Faithful Task-Agnostic Prompt Compression. \emph{arXiv preprint arXiv:2403.12968}.

{[}26{]} Yao, S., Zhao, J., Yu, D., Du, N., Shafran, I., Narasimhan, K., \& Cao, Y. (2023). ReAct: Synergizing Reasoning and Acting in Language Models. \emph{International Conference on Learning Representations (ICLR)}.

{[}27{]} Zhang, W., et al.~(2024). Coherent Microservices: Extending Cache Coherence to Distributed State Management. \emph{Proceedings of the 29th ACM Symposium on Operating Systems Principles (SOSP)}.

{[}28{]} Wang, Q., Tang, Z., Jiang, Z., Chen, N., Wang, T., \& He, B. (2025). AgentTaxo: Dissecting and Benchmarking Token Distribution of LLM Multi-Agent Systems. \emph{Published at ICLR 2025 Workshop on Foundation Models in the Wild.}

\end{document}